\documentclass[a4paper,10pt]{article}
\pdfoutput=1 
\usepackage{jheppub} 
\usepackage{epsfig}
\usepackage{caption}
\usepackage{subcaption}
\usepackage{color,soul}
\usepackage{float}
\usepackage{amsfonts}
\usepackage{url}
\usepackage{slashed}
\usepackage{lipsum}
\usepackage{mathtools}
\usepackage{physics}

\usepackage[normalem]{ulem}

\title{Time evolution of the equation of state during perturbative reheating and its impact on the inflationary tensor perturbation spectrum}
\author[a]{Avirup Ghosh}
\author[b]{and Deep Ghosh}
\affiliation[a]{ARC Centre of Excellence for Dark Matter Particle Physics, 
School of Physics, The University of Melbourne, Victoria 3010, Australia}
\affiliation[b]{Department of Physical Sciences, Indian Institute of Science Education and Research, Kolkata, Mohanpur - 741246, India}

\emailAdd{avirup.ghosh1993@gmail.com}
\emailAdd{matrideb1@gmail.com}

\abstract{
The spectrum of inflationary tensor perturbations is one of the very few  available probes of the post-inflationary reheating epoch, and it is strongly influenced by the Universe’s equation of state during this period. In the current era of precision cosmology, an accurate estimation of this primordial tensor perturbation spectrum is crucial. Unlike the conventional assumption of a constant equation of state during the perturbative reheating phase, in this work we dynamically calculate the time evolution of the reheating equation of state for different values of the inflaton decay rates including their possible time-dependence in some cases. We further investigate its impact on the spectrum of primordial tensor perturbations, focusing on two different inflationary potentials of the E-model $\alpha$-attractor class, with $n=1$ and $n=3$. Using this approach, our  results indicate that the inflationary tensor perturbations can be enhanced (or suppressed) by a factor of $\sim$ 1.5–3 compared to standard calculations assuming a constant inflaton equation of state. Such a variation arises because the evolution of the comoving horizon differs in the two approaches, causing different comoving momentum modes to re-enter the horizon at different times and undergo distinct evolutions.}

\begin{document} 
\maketitle
\flushbottom  
\section{Introduction}
\label{sec:sec1}

The early Universe underwent a phase of exponential expansion known as inflation, which is crucial for explaining its large-scale homogeneity and isotropy observed today. However, inflation must had eventually ended, transitioning the Universe into the radiation-dominated era so that the predictions of Big-Bang Nucleosynthesis (BBN) can be correctly reproduced. This transitional phase between the end of inflation and the radiation-dominated era is called the reheating era, during which the energy stored in the inflaton field was transferred presumably into the Standard Model(SM) particles, creating a hot radiation bath. Despite its importance, the reheating era remains one of the least understood epochs in the history of the early Universe due to the lack of direct observational evidence. However, this phase may have been the stage for several interesting  phenomena, such as, the dark matter production, matter-antimatter asymmetry generation~\cite{Harigaya:2014waa,Mambrini:2021zpp,Gorbunov:2010bn,Dolgov:1996qq} and hence, understanding the physics of reheating is essential to gain deeper insights into the evolution of the early Universe.

Inflation not only explains the large-scale homogeneity and isotropy of the present-day Universe but also provides a natural mechanism for the origin of primordial perturbations~\cite{Hawking:1982cz,Guth:1982ec,Starobinsky:1982ee,Bardeen:1983qw}. The scalar component of these perturbations, arising from quantum fluctuations during inflation, serve as the seeds for the formation of large-scale structures like galaxies and clusters. Moreover, inflation predicts the existence of tensor perturbations, as well, which manifest as inflationary gravitational waves (GWs). Detecting these GWs would offer strong evidence in favour of the theory of slow-roll inflation. Recent advancements in gravitational wave astronomy, such as the GW detections by LIGO~\cite{LIGOScientific:2016emj}, NANOGrav~\cite{NANOGrav:2020spf}, and other observatories, have spurred new proposals to detect GWs across a wide range of frequencies~\cite{KAGRA:2021kbb,Branchesi:2023mws,Crowder:2005nr,Seto:2001qf,Amaro-Seoane:2012vvq,Janssen:2014dka}. Such efforts could significantly enhance our understanding of the inflationary dynamics and provide deeper insights into the early Universe.

The inflationary tensor perturbations exit the cosmological horizon due to the rapid decrease in the size of the comoving horizon during inflation and later, as the Universe further evolves the comoving horizon gradually increases and these perturbations reenter the horizon. After this, such comoving modes pass through various phases, including reheating, radiation-domination, matter-domination etc. Since the gravitational waves interact very weakly with matter and radiation, they travel almost unaffected from their origin till the present day. This makes them a remarkably clean probe for studying the physics of the early Universe. In particular, the primordial GW spectrum encodes information about the inflationary era and the subsequent reheating phase. Thus, by observing the primordial GW spectrum, one can reveal the details about the reheating epoch. This potential has motivated extensive studies, as the next-generation of gravitational wave detectors provide a unique opportunity to test and refine our understanding of the physics of the early Universe (see, for example,~\cite{Kuroyanagi:2008ye,Figueroa:2019paj,Haque:2021dha,Mishra:2021wkm,Soman:2024zor}).

Specifically, the equation of state (EOS) of the Universe ($w = p_{\rm tot}/\rho_{\rm tot}$) during the reheating era plays a crucial role in shaping the primordial tensor perturbations spectra. With the advancements in the next-generation of gravitational wave (GW) detectors, there exists an increasing opportunity to accurately probe these spectra, which could enhance our understanding of the physics at play during reheating~\cite{Kuroyanagi:2008ye,Figueroa:2019paj,Haque:2021dha,Mishra:2021wkm,Soman:2024zor}. However, most of the earlier studies in this direction have considered the inflaton EOS $w_\phi$ to be a constant during reheating, but such an approximation is not always true. For example, in refs.~\cite{Antusch:2020iyq,Antusch:2021aiw,Antusch:2022mqv,Saha:2020bis} it has been shown that the time-varying inflaton EOS can stem from an era of non-perturbative particle production following the inflation. In particular, the  authors of ref.~\cite{Saha:2020bis} have parametrized the inflaton EOS, taking inputs from lattice simulations performed for specific models involving inflaton and other fields. \textit{On the contrary, in the present work we show the evolution of the inflation EOS within the perturbative reheating epoch for specific inflationary potentials, while being agnostic about the underlying model that dictates the inflaton decay rates to the visible sector.} Consequently, we demonstrate its imminent impact on the inflationary tensor perturbations for different cosmological histories stemming from different inflaton decay rates.

With these motivations, we have considered two different inflationary potentials of the E-model $\alpha$-attractor class and fixed all relevant parameters using the Cosmic Microwave Background (CMB) data of the scalar-spectral index $n_s$ and the amplitude of the scalar perturbations $A_s$. Following which, we have solved the evolution equations of the inflaton energy density and radiation energy density during the reheating era to determine the time-evolution of the reheating EOS $w_{\rm RH}$ dynamically for different values of the inflaton decay rates, including their possible time dependence in some cases.  This allows us to track the evolution of the comoving hubble radius ($1/aH$) during reheating era which impacts the spectra of primordial tensor perturbations reentering the comoving horizon during the reheating epoch. Finally, we find that our approach of using the time-varying $w_{\rm RH}$ predicts a present-day GW spectrum which is enhanced (or suppressed) compared to that predicted assuming a constant inflaton-EOS, by a factor of $\sim 1.5-3$. Although such an $\mathcal{O}(1)$ improvement can be very hard to decipher in the next-generation of gravitational wave observations due to various systematic uncertainties, we  still think it is crucial to point out such effects, as we gradually move towards an era of increasingly precise gravitational wave astronomy. 

This paper is organized as follows: In Sec.~\ref{sec:sec2}, we have 
obtained the evolutions of the reheating EOS and also the evolutions 
of the comoving horizon of the Universe during the perturbative reheating era. Sec.~\ref{sec:sec3} is devoted to determine the evolution of 
the tensor perturbations during reheating epoch and to obtain the 
impact of the time-variation of $w_\phi$ on the present-day GW 
spectrum. In Sec.~\ref{sec:sec5}, we summarize and conclude, while in Appendix.\ref{sec:AppA}, we show the slight scale variance of primordial tensor power spectrum for large wavelengths.

\section{Dynamics of inflation and post-inflationary reheating}
\label{sec:sec2}

Given a single-field potential $V(\phi)$, the inflationary dynamics is 
determined by the following slow-roll parameters:
\begin{equation}
\epsilon(k_*) = \frac{M^2_{\rm Pl}}{2}\left(\frac{V^\prime(\phi)}{V(\phi)}\right)^2\bigg|_{\phi=\phi_*}, 
\eta(k_*) = M^2_{\rm Pl}\left(\frac{V^{\prime\prime}(\phi)}{V(\phi)}\right)\bigg|_{\phi=\phi_*}, 
\end{equation} 
where $^\prime$ denotes derivative w.r.t $\phi$, $k_*$ is the CMB pivot 
scale $0.05\,{\rm Mpc}^{-1}$ and $\phi_*$ is the value of $\phi$ when 
$k_*$ exits the horizon. These slow-roll parameters are directly related 
to the CMB observables, i.e., the scalar tilt $n_s$, amplitude of density 
perturbations $A_s$ and the tensor-to-scalar ratio $r$, as follows:
\begin{equation}
n_s = 1+2\eta-6\epsilon, A_s = \frac{1}{24\pi^2\epsilon}\frac{V(\phi)}{M^4_{\rm Pl}}, r = 16\epsilon.
\end{equation}
The recent Planck data~\cite{Planck:2018vyg} suggests 
$n_s = 0.9649 \pm 0.0044$, 
$A_s = (2.105 \pm 0.029) \times 10^{-9}$ and $r < 0.1$ at the pivot 
momentum $k_* = 0.05\,{\rm Mpc}^{-1}$, which can be utilised to 
constrain the parameters of any given inflationary model. 

While most of the monomial models of chaotic inflation fail to comply with 
the recent Planck data, the $\alpha$-attractor models are one class of 
models that fit the current CMB data (see~\cite{Planck:2018jri}) quite 
well. Apart from their consistency with the CMB data these models are also  
motivated from several theoretical perspectives. Most of these 
$\alpha$-attractor models can actually arise from supergravity and 
can also explain supersymmetry breaking~\cite{Kallosh:2015lwa}. 
Additionally, the possible origin of the late time cosmological 
constant driven expansion can also be obtained within such 
models~\cite{Linder:2015qxa}. Hence, in this work, we have considered 
the general class of the E-model $\alpha$-attractor potential:
\begin{equation}
V(\phi) = \lambda\,M^4_{\rm Pl}\left(1-e^{-\sqrt{2/3\alpha}\phi/M_{\rm Pl}} \right)^{2n} \simeq \left(\frac{2}{3\alpha}\right)^n\lambda\,M^4_{\rm Pl}\,(\phi/M_{\rm Pl})^{2n},\,\,{\rm about\,\,}\phi=0 
\label{eq:infpotorig}
\end{equation}
for which,
\begin{eqnarray}
n_s &=& \frac{1 - 2(1+\frac{4n}{3\alpha})e^{-\sqrt{2/3\alpha}\phi/M_{\rm Pl}} + (1-\frac{8n^2}{3\alpha})e^{-2\sqrt{2/3\alpha}\phi/M_{\rm Pl}}}{\left(1-e^{-\sqrt{2/3\alpha}\phi/M_{\rm Pl}} \right)^{2}},\\
A_s &=& \frac{\alpha \lambda}{32 n^2 \pi^2} e^{2\sqrt{2/3\alpha}\phi/M_{\rm Pl}}\left(1-e^{-\sqrt{2/3\alpha}\phi/M_{\rm Pl}} \right)^{2(1+n)},\\
r &=& \frac{64n^2}{3\alpha} \frac{e^{-2\sqrt{2/3\alpha}\phi/M_{\rm Pl}}}{\left(1-e^{-\sqrt{2/3\alpha}\phi/M_{\rm Pl}} \right)^{2}},\\
N_k &=& \frac{1}{M^2_{\rm Pl}}\int^{\phi_{k}}_{\phi_{\rm end}} d\phi~ \frac{V(\phi)}{V'(\phi)} = \frac{1}{M^2_{\rm Pl}}\int^{\phi_{k}}_{\phi_{\rm end}} d\phi~\sqrt{\frac{3\alpha}{2}} \frac{\left(e^{\sqrt{2/3\alpha}\phi/M_{\rm Pl}}-1 \right)}{2n/M_{\rm Pl}}.
\end{eqnarray}
$N_k$ is number of $e$-foldings until the horizon exit of the pivot scale during the inflation. Assuming $\alpha = 1$ and using the central values of $n_s$, $A_s$ and 
$\epsilon(\phi_{\rm end}) = r(\phi_{\rm end})/16 = 1$ (inflation ends at 
$\phi = \phi_{\rm end}$), we obtain the parameters for two different 
potentials with $n=1$ and $n=3$, which are shown in tab.~\ref{tab:Infpot}. 
At the end of inflation, 
$\epsilon(\phi_{\rm end}) \simeq \frac{3}{2}\dot{\phi}^2/\rho_\phi|_{\phi_{\rm end}} \sim 1$, $\dot{\phi} \sim \sqrt{V(\phi_{\rm end})}$ and hence, 
$\rho_\phi \sim 3V(\phi_{\rm end})/2 \equiv \rho_{\rm end}$. We also  
assume the energy density in radiation, i.e., $\rho_R$, to be 0 at the 
end of inflation. $N_k$ is calculated with the knowledge of $\phi_k$ and $\phi_{\rm end}$. In the slow roll approximation, $\phi_{\rm end}$ becomes known and $\phi_k$ is derived using measurement of $n_s$ in the CMB scale. In particular, for quadratic inflaton potential, $N_k \approx (3+n_s)/2(1-n_s)$.

\begin{table}[htb!] 
\begin{center}
\begin{tabular}{|c|c|cc|c|c|c|}
\hline
Model & $\lambda$ & $\phi_*/M_{\rm Pl}$ & $\phi_{\rm end}/M_{\rm Pl}$ & $r$ & $N_k$ & $a_{\rm end}$ \\
& & & & & & \\\hline
& & & & & & \\
E-model ($\alpha=1, n=1$) & $1.12\times 10^{-10}$ & 5.35 & 0.94 & 0.003 & 55 & $1.47\times 10^{-29}$ \\
$\lambda\,M^4_{\rm Pl}\left(1-e^{-\sqrt{2/3}\phi/M_{\rm Pl}} \right)^2$ & & & & & & \\
$\simeq \dfrac{1}{2}\dfrac{4}{3}\lambda\,M^4_{\rm Pl}(\phi/M_{\rm Pl})^2$
& & & & & & \\
& & & & & & \\
\hline
& & & & & & \\
E-model ($\alpha=1, n=3$) & $1.14\times 10^{-10}$ & 6.67 & 1.83 & 0.003 & 56 & $5.03\times 10^{-29}$  \\
$\lambda\,M^4_{\rm Pl}\left(1-e^{-\sqrt{2/3}\phi/M_{\rm Pl}} \right)^6$ & & & & & & \\
$\simeq \dfrac{8}{27}\lambda\,M^{4}_{\rm Pl}\left(\phi/M_{\rm Pl}\right)^6$
& & & & & & \\
& & & & & & \\
\hline
\end{tabular}
\end{center}
\caption{Constraints imposed by the Planck CMB data ($n_s = 0.9649$, 
$A_{s}= 2.105 \times 10^{-9}$ at the pivot scale $k_* = 0.05\,{\rm Mpc}^{-1}$)~\cite{Planck:2018vyg} on the parameters of the chosen inflationary 
potentials are presented. The inflation field starts oscillation from $\phi = \phi_{\rm end}$. $N_k$ is the number of $e$-foldings until the horizon exit of the pivot scale during the inflation.}	
\label{tab:Infpot}
\end{table} 

\begin{figure}[htb!]
\centering
\includegraphics[width=7.5cm,height=6.2cm,angle=0]{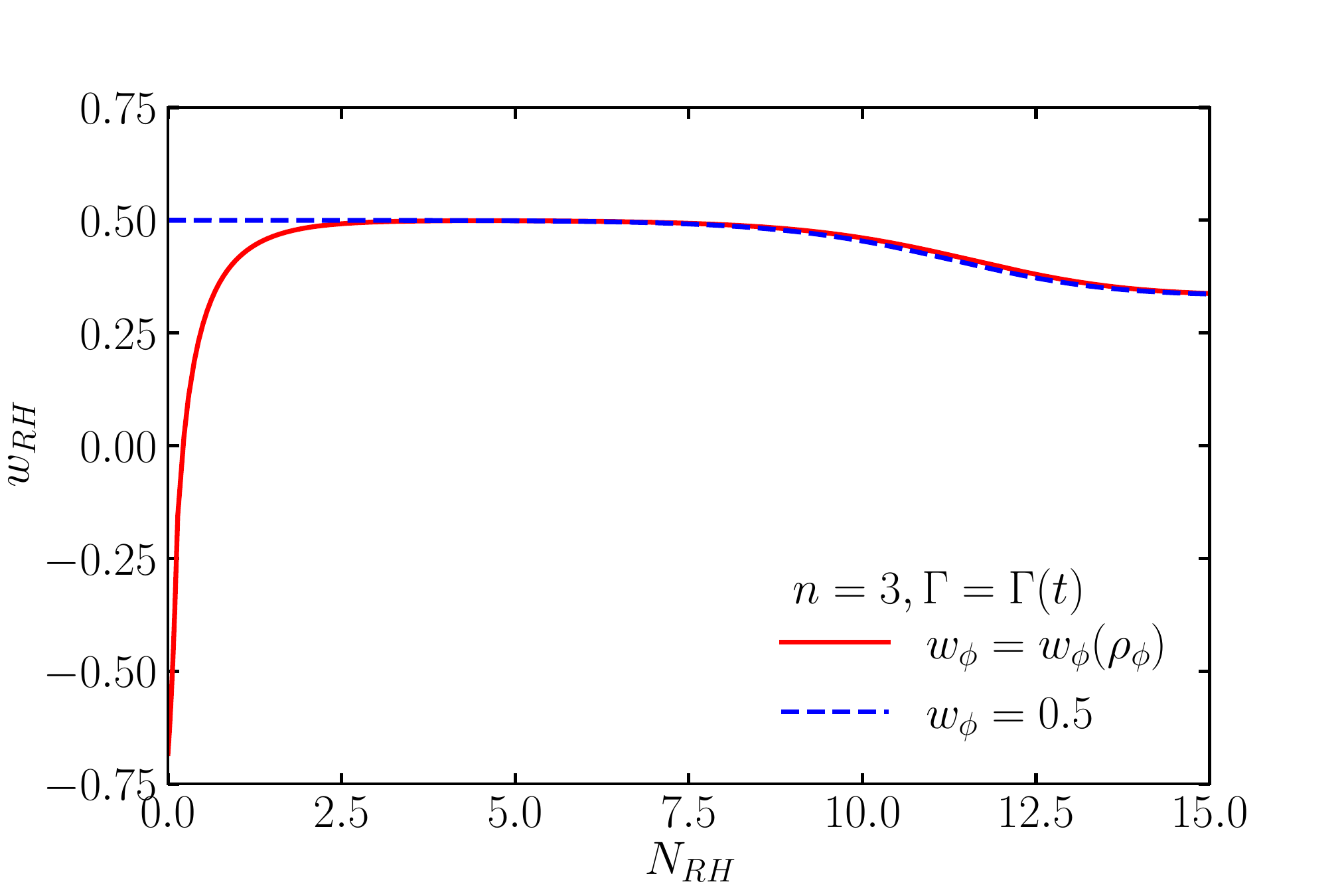}
\caption{For the $n=3$ case, we have shwon the 
differences of the evolutions of $w_{\rm RH}$ for constant inflaton-EOS (blue 
dashed) and exact EOS (red solid) assuming time-dependent inflaton decay rate. 
See text for details.}
\label{fig:EOSevol_NCD}
\end{figure}

For $\phi < \phi_{\rm end}$, the inflaton field $\phi$ starts oscillating 
about the minimum of $V(\phi)$, following the eq:
\begin{equation}
\ddot{\phi} + (3H+\Gamma)\dot{\phi}+V^{\prime}(\phi) = 0,
\label{eq:phievol}
\end{equation}
where $H$ is the hubble rate, $V^\prime(\phi) = dV(\phi)/d\phi$ and 
$\Gamma$ represents the decay rate of the inflaton condensate. The decay rate can be either time-independent or time-dependent, depending on the inflaton potential during the reheating phase. For example, for quadratic potential, the inflaton mass becomes time-independent thereby the decay rate can be parametrized by a time-independent quantity, while for the sextic potential it becomes time-dependent, stemming from the time-varying oscillation amplitude of the inflaton condensate. For details see Refs.\cite{Garcia:2020wiy,Nurmi:2015ema}. In this study we have shown the effect of time-dependent decay rate of the inflaton on the EOS of the reheating phase (see Fig.\ref{fig:EOSevol_NCD}), subsequently on the tensor power spectrum (see the right panel of Fig.\ref{fig:GWspec}) for the $\alpha-$ attractor model with  $n=3$. 

 Moreover, as mentioned in the introduction, our study points out the effect of the shape of the inflaton potential on the inflaton EOS during the perturbative reheating epoch and hence the possibility of an era of non-perturbative particle production following inflation has been ignored here. For the effect of the non-perturbative particle production on inflaton EOS we refer the reader to~\cite{Saha:2020bis}.

The energy density and the pressure density of the field $\phi$ are given by,
\begin{eqnarray}
\label{eq:phiprop}
\rho_\phi &=& \frac{\dot{\phi}^2}{2} + V(\phi),\\
p_\phi &=& \frac{\dot{\phi}^2}{2} - V(\phi),
\label{eq:phiprop1}
\end{eqnarray}
and combining eqs.\ref{eq:phiprop} and~\ref{eq:phiprop1} with eq.~\ref{eq:phievol} we obtain,
\begin{equation}
\frac{d\rho_\phi}{dt} + 3\,H\,(1+w_\phi)\,\rho_\phi = -\Gamma\,(1+w_\phi)\,\rho_\phi,
\label{eq:rhophievol}
\end{equation}
which dictates the evolution of the inflaton energy density during the 
reheating era. On the other hand, from the conservation of the 
energy-momentum tensor we get:
\begin{equation}
\frac{d\rho_{\rm tot}}{dt} + 3\,H\,\left(1+w_{\rm RH}\right) \rho_{\rm tot} = 0,
\label{eq:rhototevol}
\end{equation}
from which one obtains the evolution of the radiation energy density as follows,
\begin{equation}
\frac{d\rho_R}{dt} + 4\,H\,\rho_R =\Gamma\,(1+w_\phi)\,\rho_\phi, 
\label{eq:rhoRevol1}
\end{equation}
where the inflaton equation of state (EOS) $w_\phi = p_\phi/\rho_\phi$ and 
the reheating EOS $w_{\rm RH}$ is given by:
\begin{equation}
w_{\rm RH} = \frac{w_\phi \rho_\phi + \rho_R/3}{\rho_\phi + \rho_R},
\label{eq:wRHeq}
\end{equation} 
assuming the Universe to be a two-fluid system during the reheating era.

\begin{figure}[htb!]
\hspace*{-3mm}
\includegraphics[width=7.5cm,height=6.2cm,angle=0]{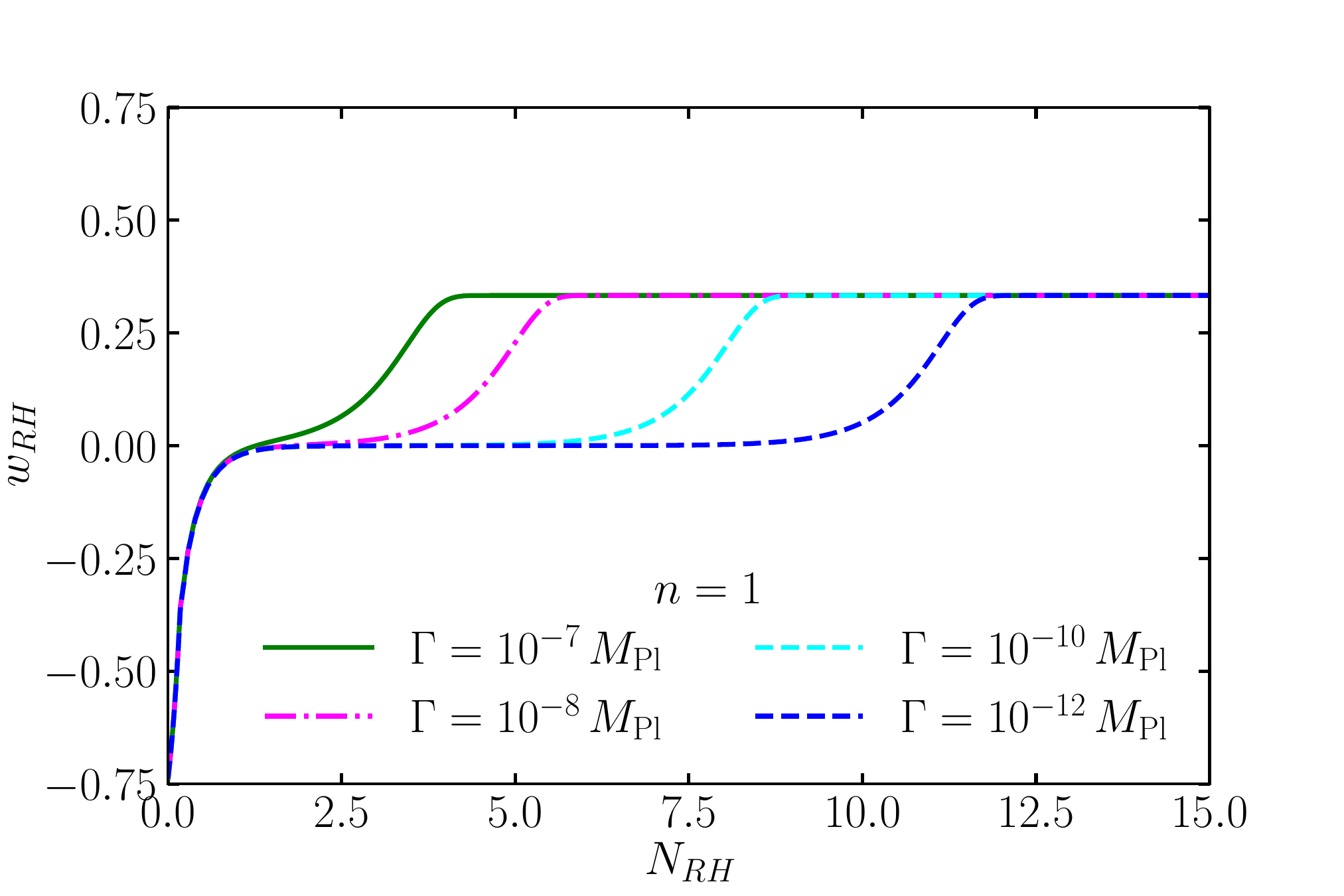}
\includegraphics[width=7.5cm,height=6.2cm,angle=0]{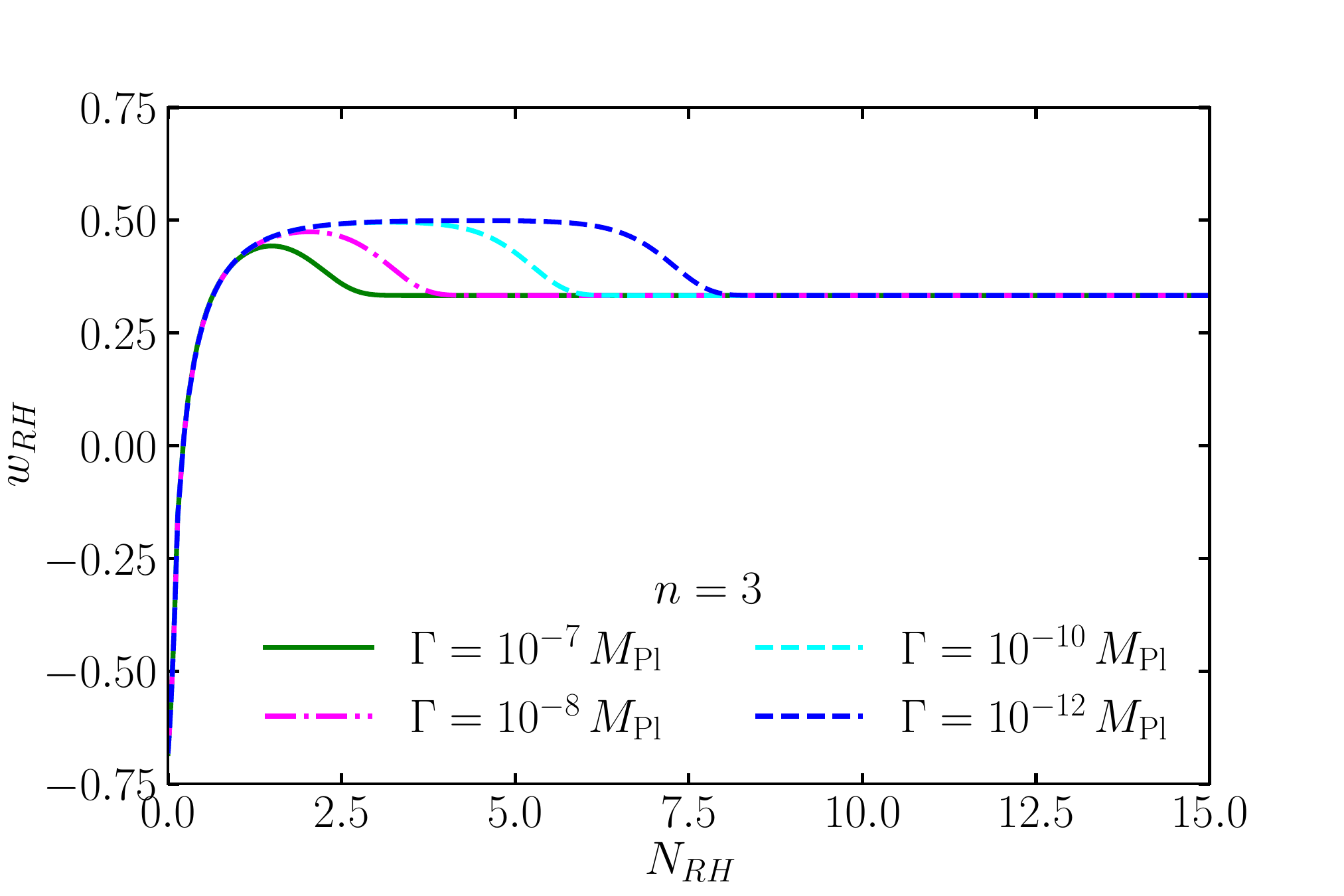}\\
\includegraphics[width=7.5cm,height=6.2cm,angle=0]{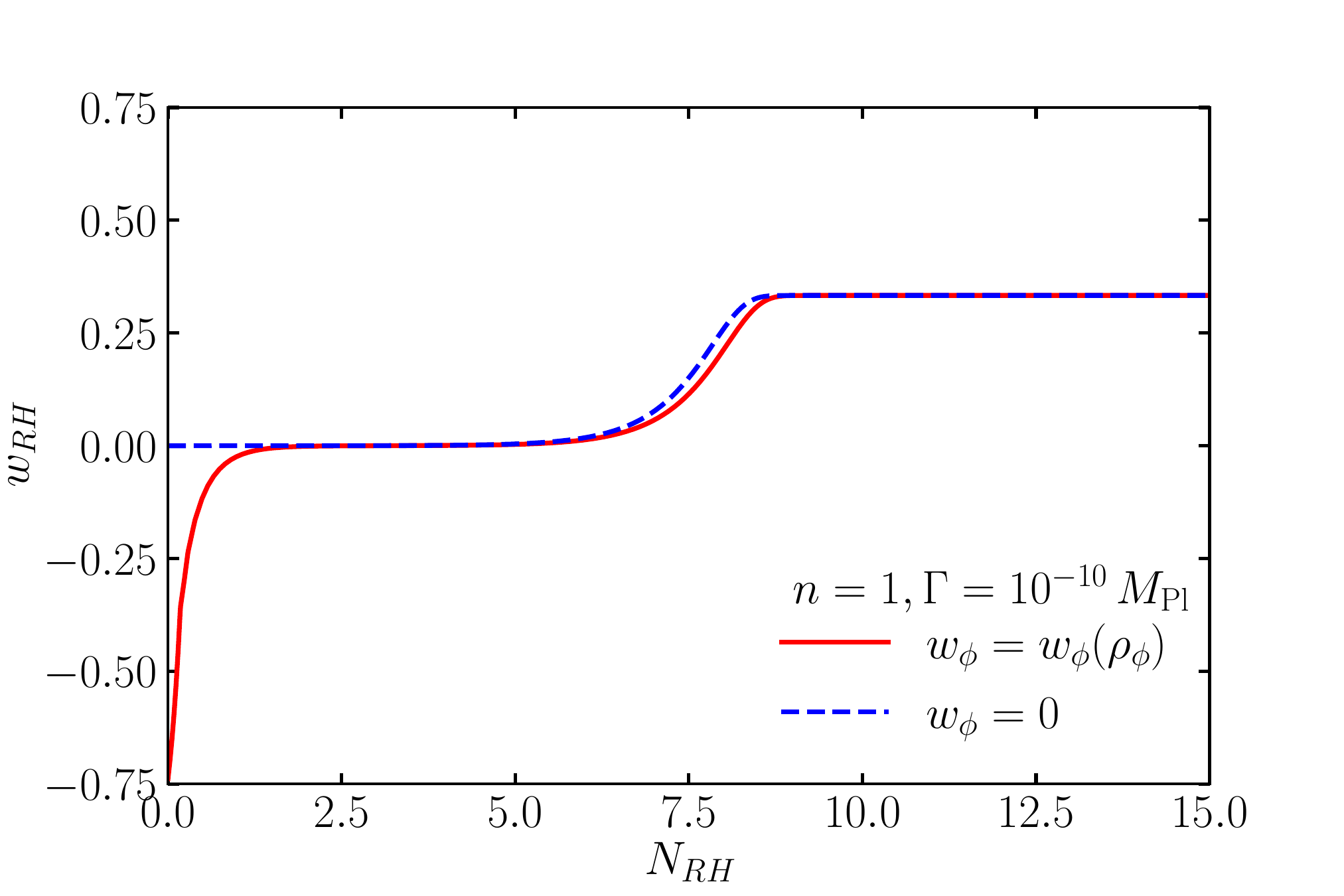}
\includegraphics[width=7.5cm,height=6.2cm,angle=0]{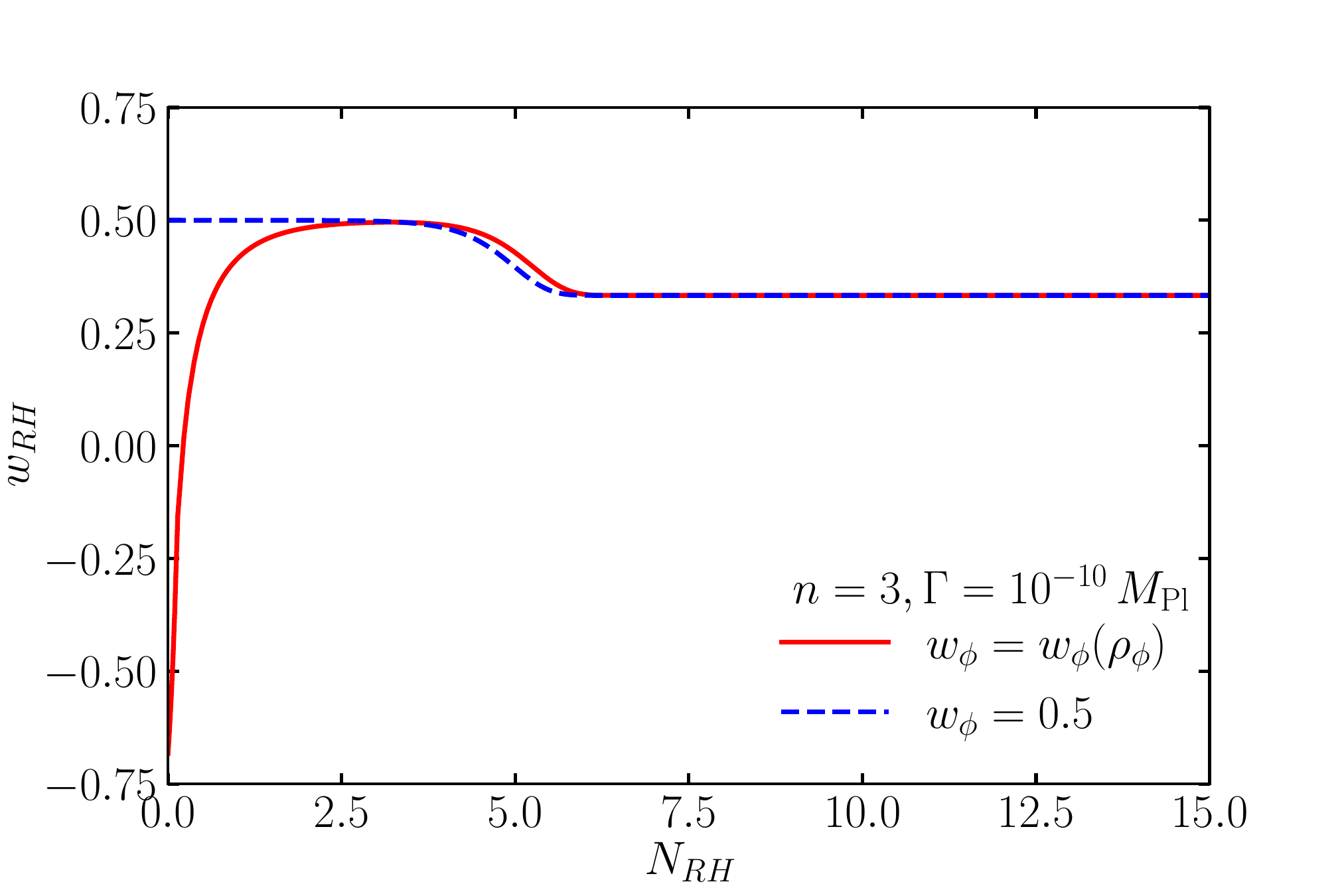}\\
\caption{{\it Top:} Evolution of the reheating EOS $w_{\rm RH}$ during  
reheating era is shown as a function of the reheating $e$-foldings 
$N_{\rm RH}$ for the E-model $\alpha$-attractor potential with 
$n=1$ (left) and $n=3$ (right). Here the four different values of the 
inflaton decay rate $\Gamma$ have been considered. {\it Bottom:} Assuming 
$\Gamma = 10^{-10}\,M_{\rm Pl}$, we have shown the evolutions of 
$w_{\rm RH}$ for constant inflaton-EOS (blue dashed) and exact EOS (red solid) 
for E-model $\alpha$-attractor potential with $n=1$ (left) and $n=3$ (right). 
See the text for details.} 
\label{fig:EOSevol}
\end{figure}

Therefore, eqs.~\ref{eq:rhophievol} and \ref{eq:rhoRevol1} together imply 
that the dynamics of reheating is affected by the value of $w_\phi$ 
and following~\cite{Turner:1983he}, $w_\phi$ can be written as,
\begin{equation}
w_\phi (\rho_\phi) = 2 \frac{\int_{\phi_{m1}}^{\phi_{m2}} d\phi \sqrt{1-V(\phi)/\rho_\phi}}{\int_{\phi_{m1}}^{\phi_{m2}} d\phi/\sqrt{1-V(\phi)/\rho_\phi}} -1,
\label{eq:wphiturner}
\end{equation}  
where $\phi_{m1,m2}$ represent the amplitudes of inflaton oscillations 
during the reheating era. The formula in eq.~\ref{eq:wphiturner} can 
capture the variations in $w_\phi$ even when the inflationary potential 
$V(\phi)$ is not absolutely symmetric around the minima of the Potential. 
For our chosen set of potentials, these are given as follows,
\begin{align} 
\phi_{m1,m2}(\rho_\phi) = \sqrt{\dfrac{3}{2}} M_{\rm Pl} \ln \left[ \dfrac{1}{1 \pm  \left(\dfrac{\rho_\phi}{\lambda M^4_{\rm Pl}} \right)^{1/2n}}\right]. 
\end{align}
Using eq.~\ref{eq:wphiturner} we first obtain the inflaton EOS 
$w_\phi(\rho_\phi)$ as a function of the inflaton energy density 
$\rho_\phi$ and then use it to solve the eqs.~\ref{eq:rhophievol} 
and~\ref{eq:rhoRevol1}. Hereafter, we call this approach to be 
`actual EOS' approach. The resulting evolutions of $\rho_\phi$ 
and $\rho_R$ are then used to obtain the reheating EOS $w_{\rm RH}$ 
from eq.~\ref{eq:wRHeq}. In the top panel of Fig.~\ref{fig:EOSevol} 
we have shown the evolutions of $w_{\rm RH}$ for $n=1$ (left) and 
$n=3$ (right) for several values of the inflaton decay rate $\Gamma$. 

Note that, if one approximates the potentials by the leading 
terms of their Taylor-series expansions about their corresponding 
minima, eq.~\ref{eq:wphiturner} gives $w_\phi = 0$ for $n=1$ and 
$w_\phi = 0.5$ for $n=3$, respectively. We also solve
eqs.~\ref{eq:rhophievol} and ~\ref{eq:rhoRevol1} using these constant 
values of $w_\phi$ and hereafter refer this approach as the `constant inflaton-EOS' approach. In the bottom panel of Fig.~\ref{fig:EOSevol} we have 
compared the evolutions of $w_{\rm RH}$ as obtained in the actual EOS 
case (red solid) and in the constant inflaton-EOS approach (blue dashed) for 
$n=1$ (left) and $n=3$ (right) assuming $\Gamma = 10^{-10}\,M_{\rm Pl}$.  

 In addition, as mentioned earlier we have also considered a 
case of time-dependent decay rate for the inlfaton in $n=3$ model. Assuming 
the inflaton decays into fermions dominantly we use~\cite{Garcia:2020wiy}:
\begin{eqnarray}
\Gamma(t) &=& M_{\rm Pl}\frac{y^2}{8\pi}\sqrt{\left(\frac{2}{3\alpha}\right)^{n}2n(2n-1)\lambda}\alpha_y\,\frac{\omega}{m_\phi}\left(\frac{\rho_\phi}{M_{\rm Pl}}\right)^{\frac{n-1}{2n}}, 
\end{eqnarray}
which for $n=3$ is given by, $\Gamma(t) = 4.61\times 10^{-7}\,M_{\rm Pl}y^2\,\left(\frac{\rho_\phi}{M_{\rm Pl}}\right)^{\frac{1}{3}}$~\cite{Garcia:2020wiy} where $y$ represents the Yukawa coupling between the inflaton 
$\phi$ and the daughter fermions. Choosing $y=1$ and putting the resulting
$\Gamma=\Gamma(t)$ in eqs.~\ref{eq:rhophievol} and \ref{eq:rhoRevol1} we 
track the evolution of $w_{\rm RH}$ for both constant $w_\phi=0.5$ and 
actual $w_\phi$ (obtained from eq.~\ref{eq:wphiturner}), which are 
presented in Fig.~\ref{fig:EOSevol_NCD}.

\begin{figure}[htb!]
\centering
\includegraphics[width=7.5cm,height=6.2cm]{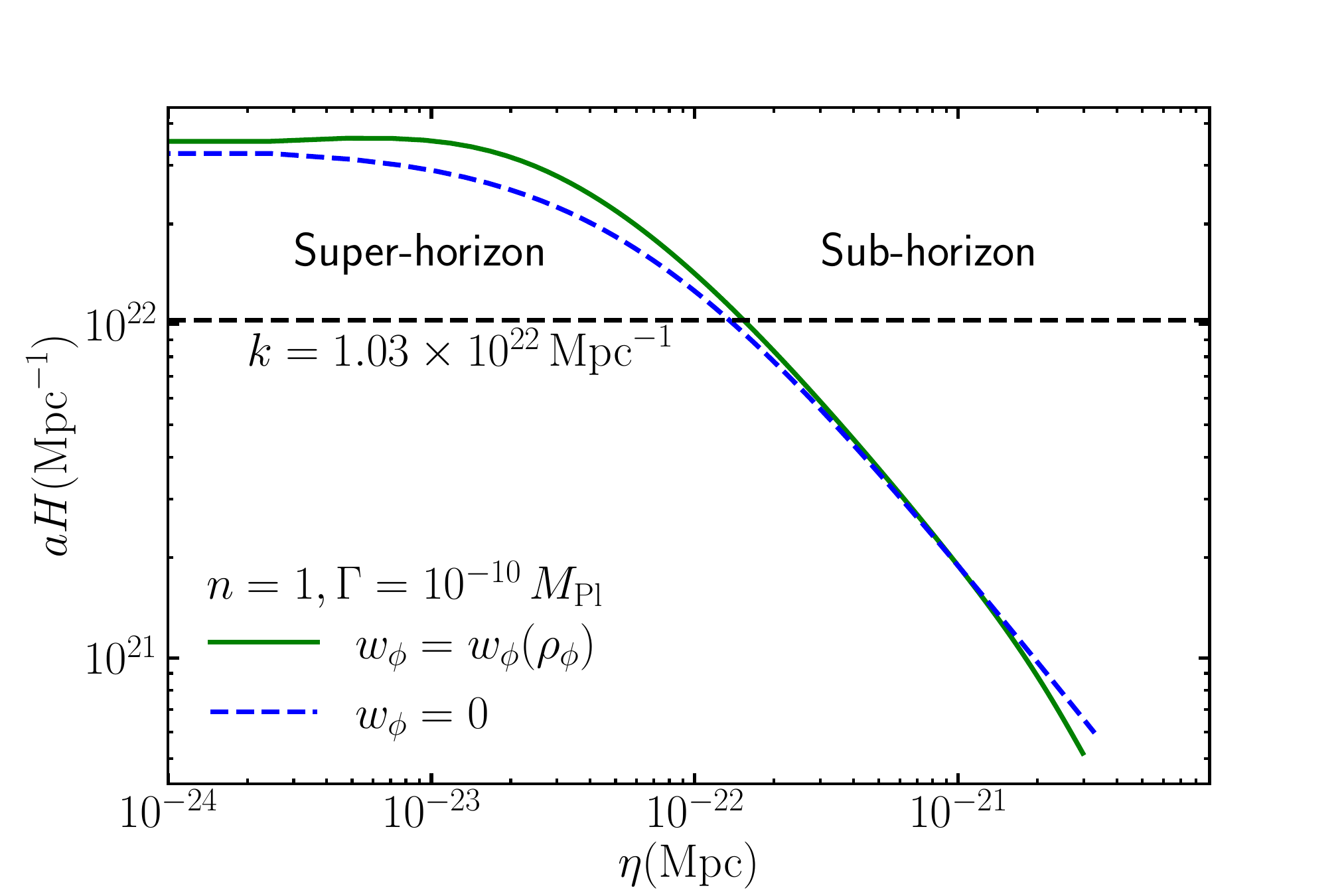}
\includegraphics[width=7.5cm,height=6.2cm]{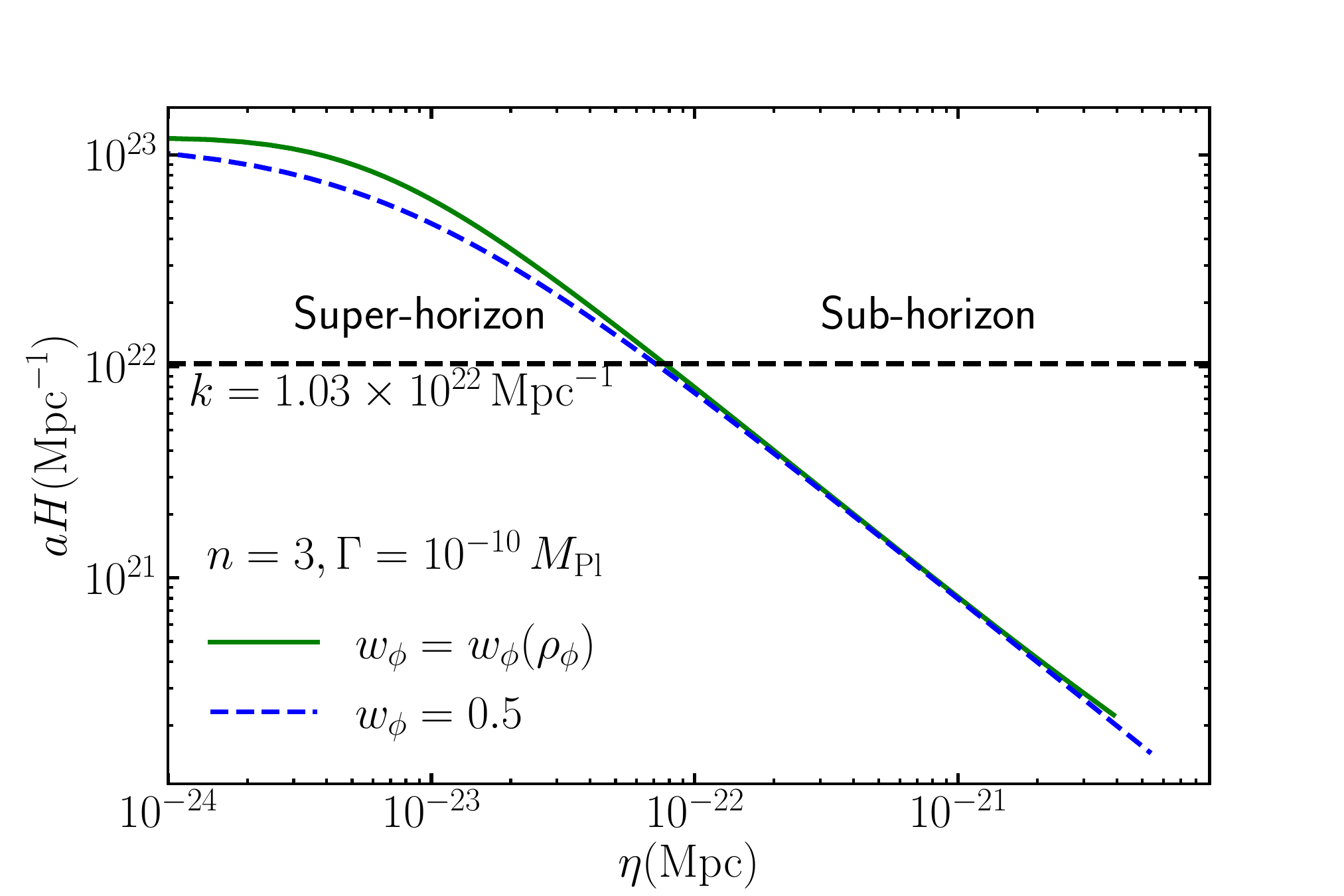}
\caption{
For the E-model $\alpha$-attractor potential with $n=1$ (left) 
and $n=3$ (right), the evolution of $a\,H$ have been shown as a function 
of the comoving time $\eta$. In each case, the blue dashed line shows the 
result obtained assuming a constant inflaton-EOS and the green solid line shows 
the evolution when the time-varying EOS is considered. Here 
$\Gamma = 10^{-10}\,M_{\rm Pl}$ is assumed. See the text for 
details.}
\label{fig:cohor}
\end{figure}

As will be evident later, another quantity of interest for our 
study is the comoving horizon ($1/aH$) of the Universe during the 
reheating era. The inverse of the comoving horizon in the constant 
EOS approach is given by,
\begin{equation}
a\,H(N) = \frac{a_{\rm end}\sqrt{\rho_{\rm end}}}{\sqrt{3}M_{\rm Pl}}e^{-N(1+3w_\phi)/2}.
\label{eq:aHanalytic}
\end{equation}
On the other hand, this same quantity for the actual EOS approach is 
obtained as follows,
\begin{equation}
a\,H(N) = \frac{a_{\rm end}\sqrt{\rho_{\rm end}}}{\sqrt{3}M_{\rm Pl}}e^{N-3\,\int_0^N\,dN^\prime\,(1+w_{\rm RH}(N^\prime))}.
\label{eq:aHnumerics}
\end{equation}
We have defined the end of reheating by the condition 
$\rho_\phi = \rho_R$. The differences of these quantities are 
evident from Fig.~\ref{fig:cohor} where we have shown $aH$ as 
a function of the comoving time $\eta = \int da/a^2H$ for $n=1$ (left) 
and $n=3$ (right), respectively. In both the cases, we have considered 
$\Gamma = 10^{-10}\,M_{\rm Pl}$ and have represented the results 
obtained for the constant inflaton-EOS approach by blue dashed lines while that 
for the actual EOS case by the green solid lines. The black dashed line 
represents a particular comoving momentum 
$k=1.03 \times 10^{22}\,{\rm Mpc}^{-1}$ which enters the horizon during the 
reheating era and hence will carry the imprints of the physics occurring 
during the reheating era.

\section{Propagation of tensor perturbations during reheating era}
\label{sec:sec3}

As mentioned earlier, inflationary tensor perturbations (or gravitational 
waves) can provide new avenues to unveil the dynamics 
of reheating era thanks to the current advancements in the field 
of GW astronomy~\cite{Kuroyanagi:2008ye,Figueroa:2019paj,Haque:2021dha,Mishra:2021wkm,Soman:2024zor}. Therefore, in this section, 
we seek to investigate how does the reheating eq. of states obtained 
in the cosntant EOS approach and in the actual EOS approach affect the  
inflationary tensor perturbation spectra. 

During inflation, the comoving hubble radius ($1/aH$) decreases and 
any particular comoving momentum mode $k$ of tensor perturbations 
produced during inflation become super-horizon when $k < a\,H$. 
After inflation, the comoving hubble radius ($1/aH$) increases and 
some of the super-horizon modes which earlier exited, reenter the 
horizon. After reentering the horizon such modes experience 
the effects of the evolution of the Universe and propagate following,
\begin{eqnarray}
\ddot{h}_{k} + \frac{3\dot{a}}{a}\dot{h}_{k} + \frac{k^2}{a^2}h_{k} = 0,
\label{eq:GWevol1}
\end{eqnarray}
in absence of any additional source term. In eq.~\ref{eq:GWevol1}, 
$h_{k}$ represents the fourier transform of the amplitude of mode 
$k$ and the polarization index is suppressed for brevity. When 
written in terms of the comoving time $\eta$, eq.~\ref{eq:GWevol1} 
becomes,
\begin{eqnarray}
h^{\prime\prime}_{k} + 2\,a H\,h^{\prime}_{k} + k^2\,h_{k} = 0,
\label{eq:GWevol2}
\end{eqnarray} 
where $^\prime$ denotes the derivative w.r.t $\eta$. 
Eq.~\ref{eq:GWevol2} can be further rewritten as:
\begin{eqnarray}
\frac{d^2h_{k}}{d(k\eta)^2} + 2\,\left(\frac{a H}{k}\right)\,\frac{dh_{k}}{d(k\eta)} + h_{k} = 0,
\label{eq:GWevol3}
\end{eqnarray} 
which is simply the equation of motion of a damped harmonic oscillator 
with time-dependent damping term, $a H/k$. Therefore, for any given comoving 
momentum $k \ll a H$ eq.~\ref{eq:GWevol3} has the solution 
$h_k = {\rm const}$ and the amplitude of the GW spectrum remains unchanged. 
On the other hand, for any given comoving momentum $k \gtrsim a H$, the 
amplitude of GW spectrum damps as the Universe evolves. To obtain the exact 
evolution of the GW spectra for constant inflaton-EOS and actual EOS approaches we 
solve eq.~\ref{eq:GWevol3} numerically, using the $a H(\eta)$ 
obtained in sec.~\ref{sec:sec2}. 

\begin{figure}[htb!]
\centering
\includegraphics[width=7.5cm,height=6.2cm]{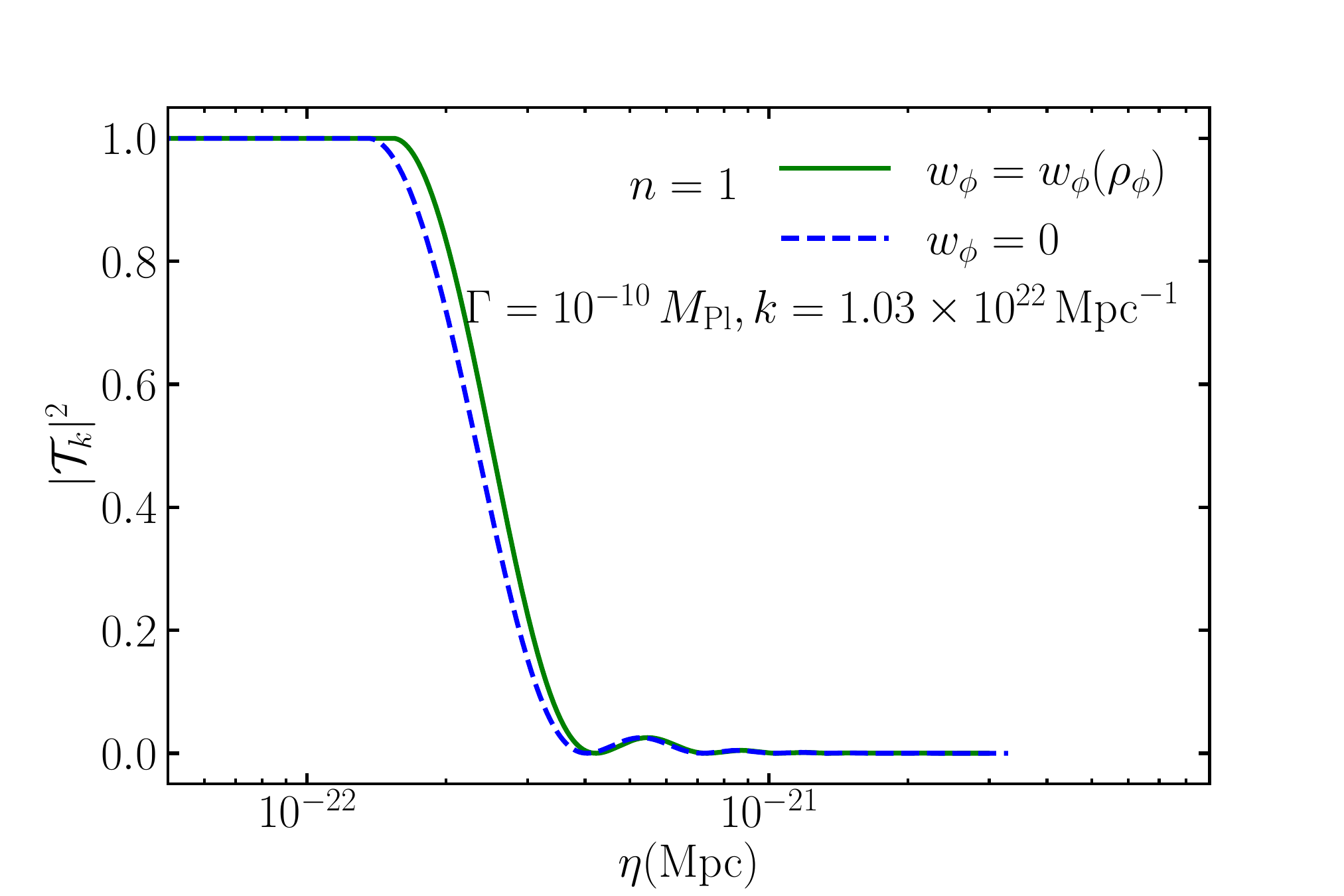}
\includegraphics[width=7.5cm,height=6.2cm]{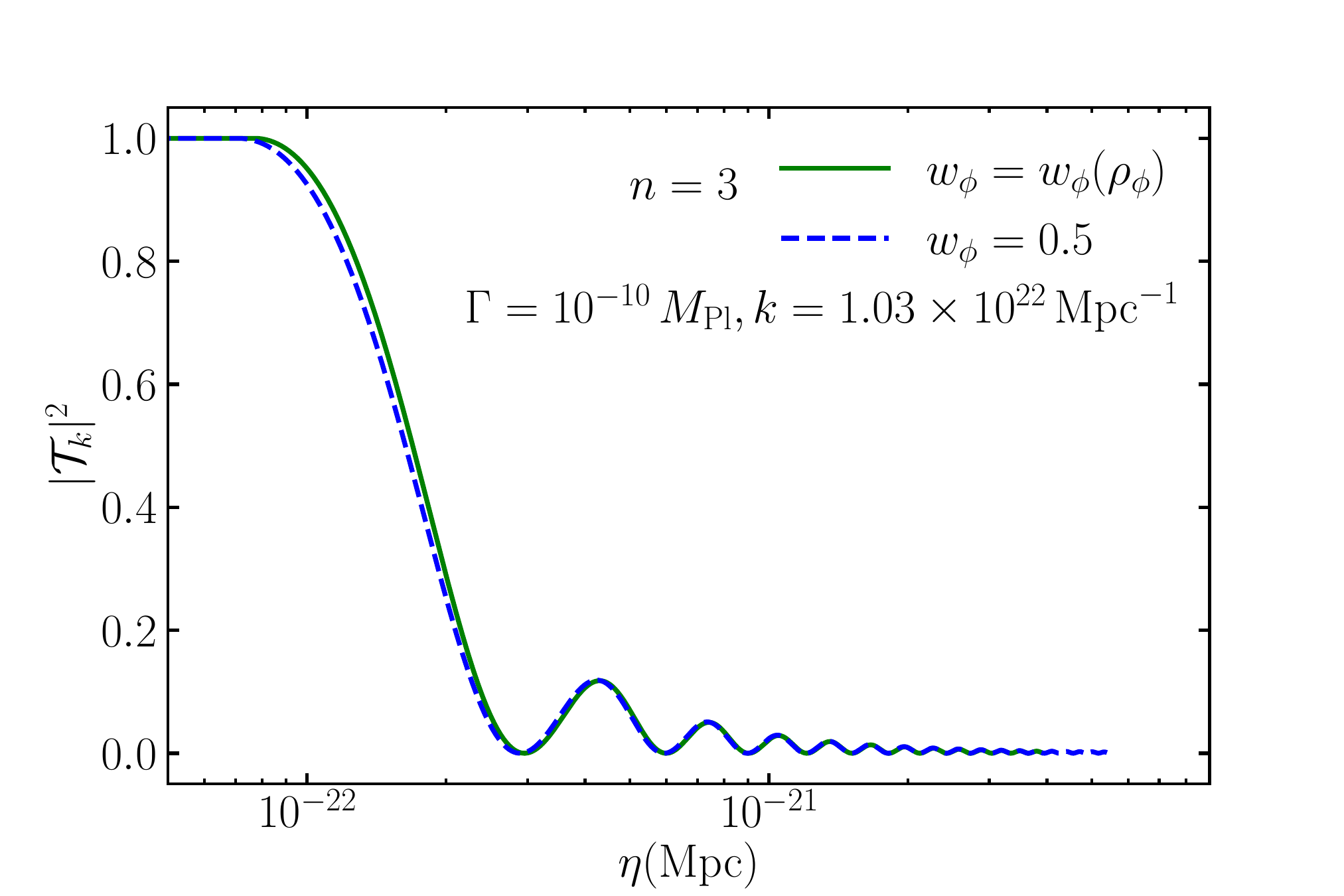}
\caption{
The evolution of a particular comoving momentum mode
$k=1.03\times 10^{22}\,{\rm Mpc}^{-1}$ of tensor perturbations, 
that enters the horizon during reheating, is shown for $n=1$ (left) 
and $n=3$ (right) assuming constant inflaton-EOS (blue dashed line) 
and actual EOS (green solid line), respectively. Here 
$\Gamma = 10^{-10}\,M_{\rm Pl}$ is assumed. See the text 
for details.}
\label{fig:GWeff}
\end{figure}

To this end, we define the transfer function and its derivative as, 
$\mathcal{T}_k(\eta) = h_{\vec{k}}(\eta)/h^{\rm prim}_{\vec{k}}$ and 
$\mathcal{T}^\prime_k(\eta) = h^\prime_{\vec{k}}(\eta)/h^{\rm prim}_{\vec{k}}$, respectively, with $h^{\rm prim}_k$ representing the 
amplitude of mode $k$ at the horizon exit during inflation. Evolution 
of $|\mathcal{T}_k(\eta)|^2$ for the mode 
$k = 1.03 \times 10^{22}\,{\rm Mpc}^{-1}$ during perturbative reheating 
era are shown in Fig.~\ref{fig:GWeff} for $n=1$ (left) and $n=3$ (right), 
assuming, $\Gamma = 10^{-10}\,M_{\rm Pl}$. Here, the blue dashed 
lines show the result for constant inflaton-EOS and the green solid lines 
are obtained considering the actual evolution of $w_{\rm RH}$. 
Clearly, the differences between the transfer functions obtained
in the two different approaches stem from when a particular 
comoving momentum mode reenters the horizon and also from the instant 
of time signifying the end of the reheating era. As we shall see 
later, these can have substantial impact on the present-day spectrum 
of inflationary tensor perturbations.

After the end of reheating, the standard radiation domination takes over 
and the amplitude of tensor perturbations evolve as:
\begin{equation}
h^{\rm RD}_k(\eta) = \sqrt{\frac{2}{\pi}}\,\left[C^k_1\,j_0(k\eta)+C^k_2\,y_0(k\eta)\right]\,h^{\rm prim}_k.
\label{eq:RDevolGW}
\end{equation}
We choose the coefficients $C^k_{1,2}$ such that at 
$\eta = \eta_{\rm RH}$ our solutions for $\mathcal{T}_k(\eta_{\rm RH})$ 
matches with eq.~\ref{eq:RDevolGW}:
\begin{eqnarray}
C^k_1 &=& \sqrt{\frac{\pi}{2}}\left(\frac{y_1(k\eta_{\rm RH})\mathcal{T}_k(\eta_{\rm RH}) - y_0(k\eta_{\rm RH})\mathcal{T}^{\prime}_k(\eta_{\rm RH})}{j_0(k\eta_{\rm RH})y_1(k\eta_{\rm RH})-y_0(k\eta_{\rm RH})j_1(k\eta_{\rm RH})}\right),\\
C^k_2 &=& -\sqrt{\frac{\pi}{2}} \left(\frac{j_1(k\eta_{\rm RH})\mathcal{T}_k(\eta_{\rm RH}) - j_0(k\eta_{\rm RH})\mathcal{T}^{\prime}_k(\eta_{\rm RH})}{j_0(k\eta_{\rm RH})y_1(k\eta_{\rm RH})-y_0(k\eta_{\rm RH})j_1(k\eta_{\rm RH})}\right).
\end{eqnarray} 
Similarly, at the matter-radiation equality, 
$\eta = \eta_{\rm eq}$, it is ensured that the solutions smoothly 
transit to the solutions in matter-dominated era, given by:
\begin{equation}
h^{\rm MD}_k(\eta) = \sqrt{\frac{2}{\pi}}\,\frac{\eta_{\rm eq}}{\eta}\left[D^k_1\,j_1(k\eta)+D^k_2\,y_1(k\eta)\right]\,h^{\rm prim}_k,
\label{eq:MDevolGW}
\end{equation}  
with 
\begin{eqnarray}
D^k_1 &=& \left[C^k_1\,\left( \frac{3}{2k\eta_{\rm eq}}- \frac{\cos(2k\eta_{\rm eq})}{2k\eta_{\rm eq}} + \frac{\sin(2k\eta_{\rm eq})}{(k\eta_{\rm eq})^2}\right) + C^k_2\,\left(1-\frac{1}{(k\eta_{\rm eq})^2}-\frac{\cos(2k\eta_{\rm eq})}{(k\eta_{\rm eq})^2} - \frac{\sin(2k\eta_{\rm eq})}{2k\eta_{\rm eq}}\right)  \right],\nonumber\\\\
D^k_2 &=& \left[C^k_1\,\left(-1+\frac{1}{(k\eta_{\rm eq})^2}-\frac{\cos(2k\eta_{\rm eq})}{(k\eta_{\rm eq})^2} - \frac{\sin(2k\eta_{\rm eq})}{2k\eta_{\rm eq}}\right) + C^k_2\,\left(\frac{3}{2k\eta_{\rm eq}}+ \frac{\cos(2k\eta_{\rm eq})}{2k\eta_{\rm eq}} - \frac{\sin(2k\eta_{\rm eq})}{(k\eta_{\rm eq})^2}\right)  \right].\nonumber\\ 
\end{eqnarray}

\begin{figure}[t!]
\includegraphics[width=7.5cm,height=6.2cm]{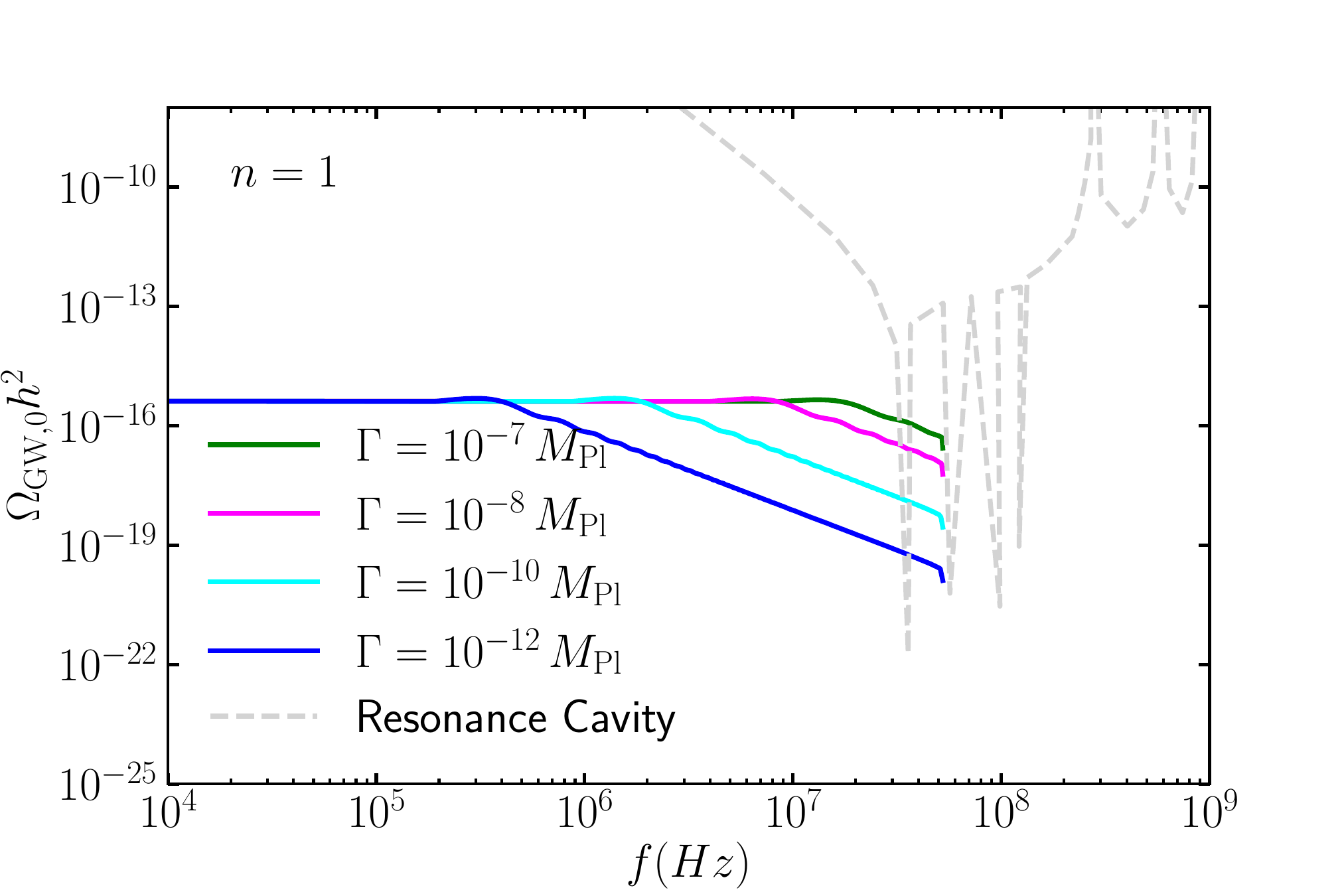}
\includegraphics[width=7.5cm,height=6.2cm]{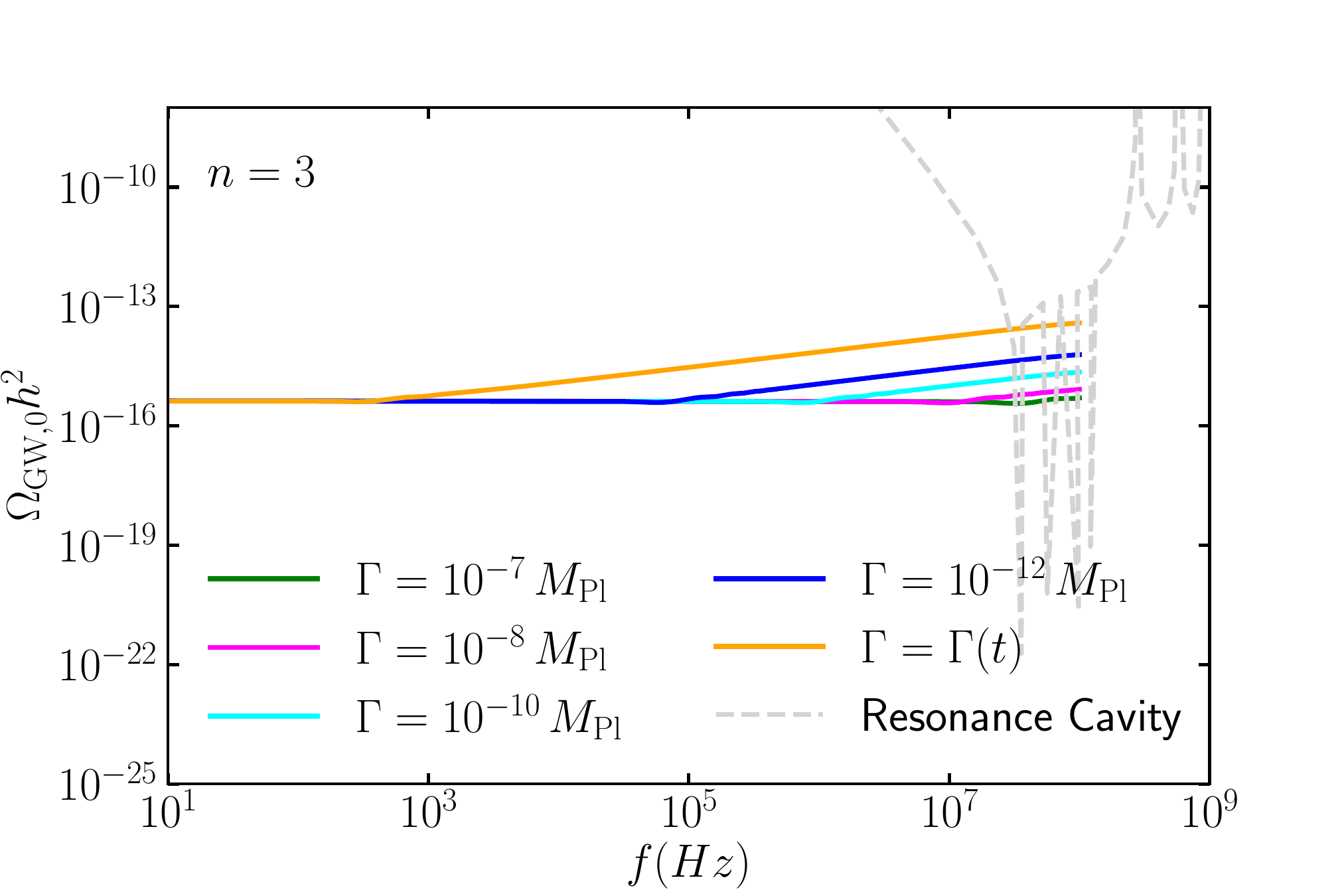}\\
\includegraphics[width=7.5cm,height=6.2cm]{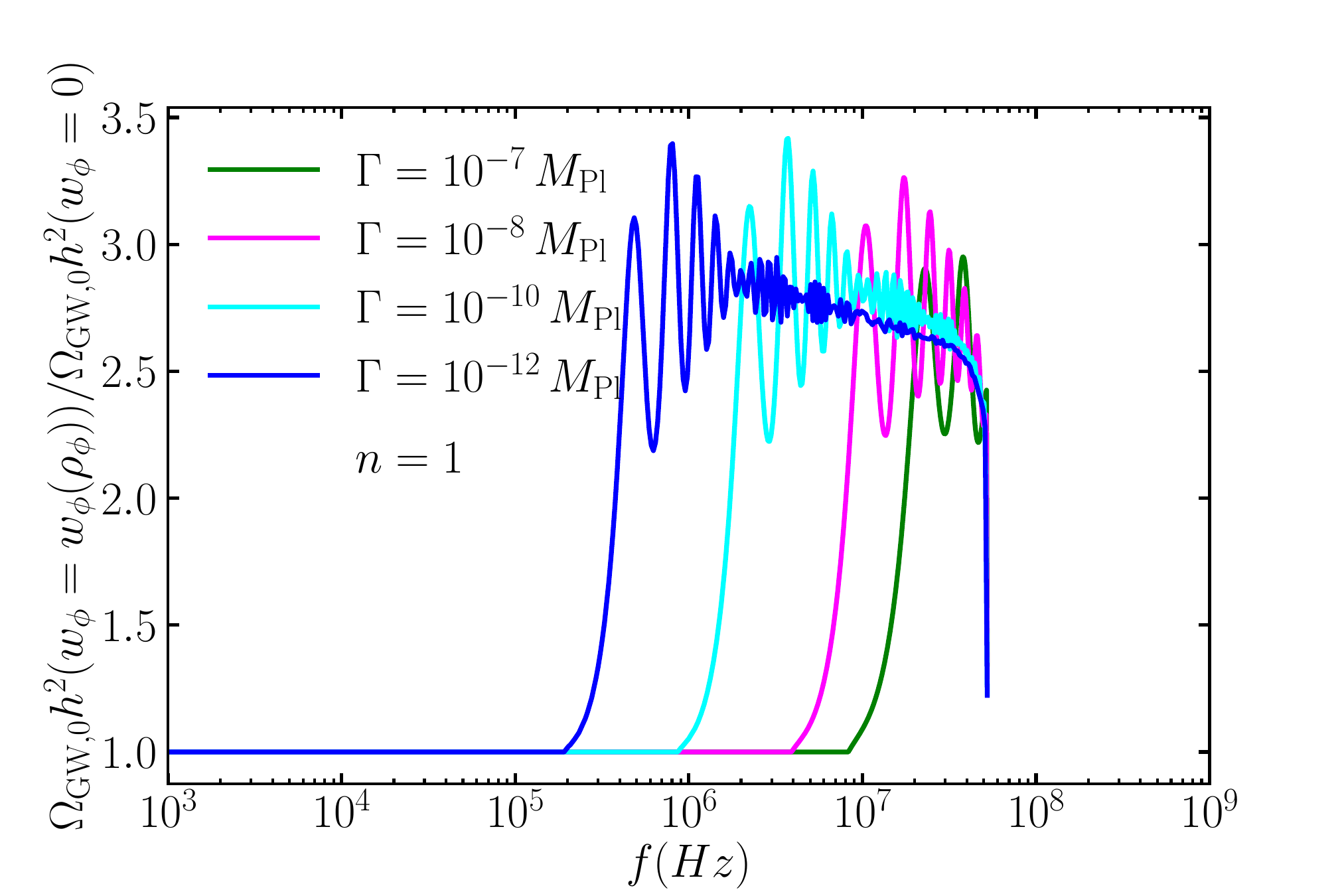}
\includegraphics[width=7.5cm,height=6.2cm]{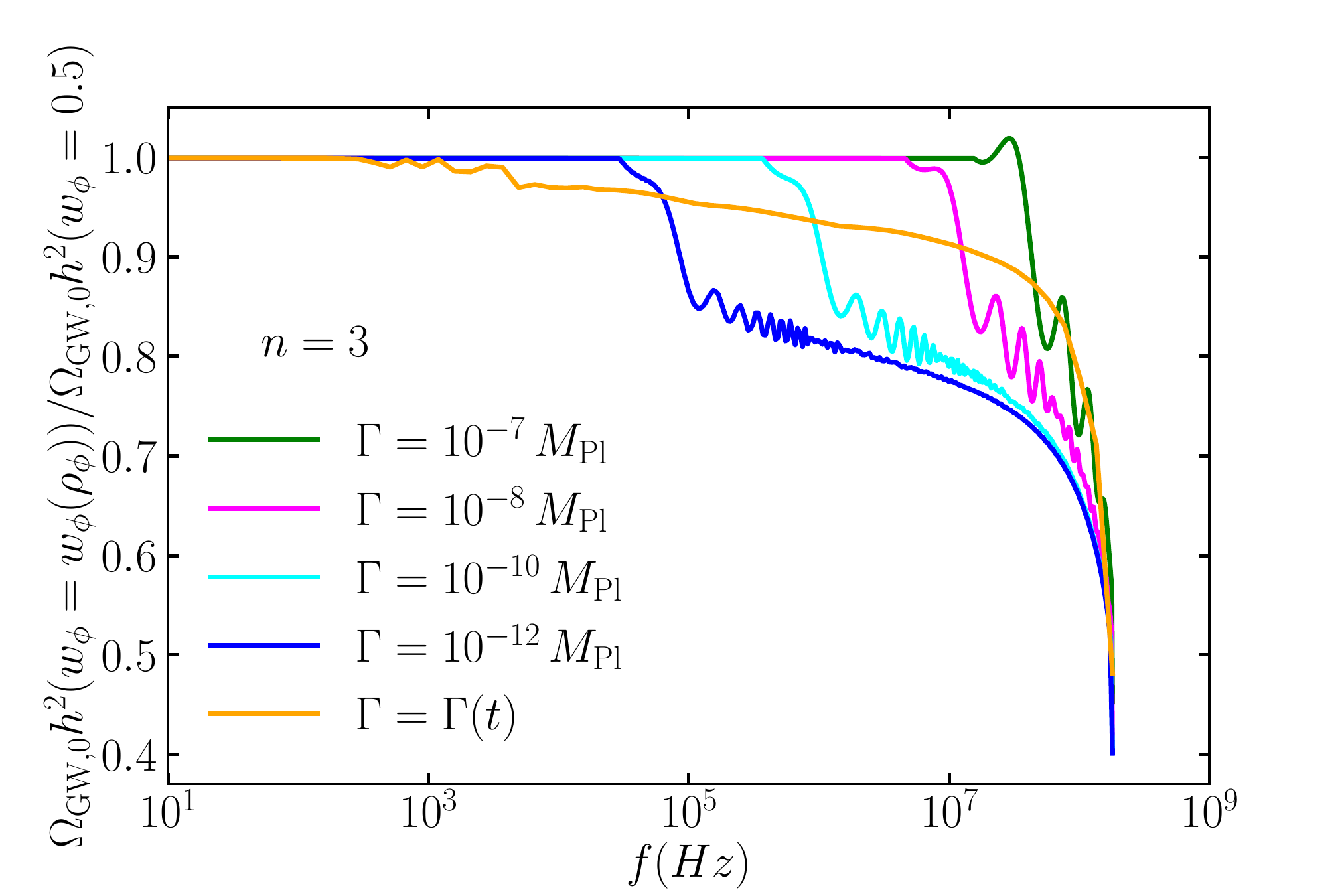}
\caption{{\it Top:} The present day spectrum of inflationary gravitational  
waves (GW) for the E-model $\alpha$-attractor potential with $n=1$ (left)  
and $n=3$ (right) are shown for different values of inflaton decay rate  
considering actual EOS. The gray dashed curves show the projected 
sensitivity of the proposed resonance cavity observations. 
{\it Bottom:} The ratio of the GW energy density considering actual EOS to 
that obtained for constant inflaton-EOS are shown for $n=1$ (left) and 
$n=3$ (right), for different choices of $\Gamma$. See the text for 
details.}
\label{fig:GWspec}
\end{figure}

Ignoring the small period of dark energy domination in the 
late Universe, the amplitude of tensor perturbations today 
(i.e., at $\eta = \eta_0$), is given by,
\begin{equation}
h_k(\eta_0) = \sqrt{\frac{2}{\pi}}\,\frac{\eta_{\rm eq}}{\eta_0}\left[D^k_1\,j_1(k\eta_0)+D^k_2\,y_1(k\eta_0)\right]\,h^{\rm prim}_k.
\label{eq:gravamp0}
\end{equation}
Now using eq.~\ref{eq:gravamp0} we obtain the GW spectral energy density 
today,
\begin{equation}
\frac{d\rho_{\rm GW}}{d\ln\,k} = \frac{M^2_{\rm Pl}}{4\,a^2(\eta_0)}\,\sum_{+,\times}\frac{2k^3}{2\pi^2}\,\langle|h^\prime_k(\eta_0)|^2\rangle.
\label{eq:gravspec0}
\end{equation} 
Using eq.~\ref{eq:gravamp0} and introducing the definition of 
primordial power spectrum $\Delta^{2}_{k,{\rm inf}} = \underset{+,\times}{\sum}\frac{2k^3}{2\pi^2}\,\langle|h^{\rm prim}_k|^2\rangle$~\footnote{Note 
that, the primordial tensor power-spectrum $\Delta^{2}_{k,{\rm inf}} \equiv P_T$ is strictly scale-invariant for $k\approx k_*$, but can varies 
slightly for large-$k$, as shown in~\cite{Garcia:2024rwg,Haque:2021dha}. 
The effect of such variation is presented in Appendix~\ref{sec:AppA}.}, 
eq.~\ref{eq:gravspec0} can be re-expressed as, 
\begin{equation}
\frac{d\rho_{\rm GW}}{d\ln\,k} = \frac{M^2_{\rm Pl}\,k^2}{4\,a^2(\eta_0)}\,\Delta^{2}_{k,{\rm inf}}\,\left\langle\bigg|\frac{dh_k(\eta)/h^{\rm prim}_k}{d(k\eta)}\bigg|^2_{\eta=\eta_0}\right\rangle,
\label{eq:gravspec0A}
\end{equation}  
where $\langle ... \rangle$ represents average over multiple oscillations. 
The relative energy density in GW today is,
\begin{equation}
\Omega^0_{\rm GW} = \frac{1}{\rho_c}\frac{d\rho_{\rm GW}}{d\ln\,k} = \frac{k^2}{12\,a^2(\eta_0)\,H^2(\eta_0)}\,\Delta^{2}_{k,{\rm inf}}\,\left\langle\bigg|\frac{dh_k(\eta)/h^{\rm prim}_k}{d(k\eta)}\bigg|^2_{\eta=\eta_0}\right\rangle,
\label{eq:gravspec0B}
\end{equation}
where the comoving momentum $k$ has an one-to-one correspondence with the 
frequency of GW signal $f=k/2\pi a_0$. In our calculation of 
$\Omega^0_{\rm GW}$ (eq.~\ref{eq:gravspec0B}) we have used 
$\eta_{\rm eq} = a_{\rm eq}/H_0\,\sqrt{\Omega_{R,0}}$, 
$\eta_0 = \eta_{\rm eq} \sqrt{\Omega_{M,0}/\Omega_{R,0}}$ and 
$H_0 = 100h\,{\rm km}\,{\rm s}^{-1}\,{\rm Mpc}^{-1}$ with 
$h=0.68$~\cite{Planck:2018vyg}.

The quantity $\Omega^0_{\rm GW}h^2$ for the E-model $\alpha$-attractor 
potential with $n=1$ and $n=3$ are shown for different values of $\Gamma$ 
in the top left and top right panels of Fig.~\ref{fig:GWspec}, 
respectively. Clearly, as obtained in the constant inflaton-EOS case, the 
GW spectrum is red-tilted for $n=1$ and blue-tilted for $n=3$. The 
ratio of $\Omega^0_{\rm GW}h^2$ for actual EOS to $\Omega^0_{\rm GW}h^2$ 
for constant inflaton-EOS are presented in the bottom panel of 
Fig.~\ref{fig:GWspec} for several values of $\Gamma$. Here, it is 
clear that for $n=1$ case (left) the GW spectrum is enhanced by a factor 
of $\sim 3$ while for $n=3$ (right) it is suppressed by a factor 
of $\sim 1.5$.  Note that, this enhancement or suppression in 
the GW spectrum is independent of whether $\Gamma$ is chosen to be a 
constant or it has a specified time dependence. This is evident from our 
choice of $\Gamma=\Gamma(t)$ in the $n=3$ model the results for which are
shown by orange lines in both top-right and bottom-right panels of 
Fig.~\ref{fig:GWspec}. The gray dashed curves show the sensitivity of the 
recently proposed resonance cavity experiments~\cite{Herman:2020wao,Herman:2022fau} which may be effective in deciphering the high-frequency tail of 
the inflationary tensor perturbation spectrum.

\section{Summary and conclusion}
\label{sec:sec5}

The early universe transits from an inflationary phase to the radiation-dominated phase through an era of reheating, facilitated by the dissipation of the inflaton field to the SM degrees of freedom. One of the well-known mechanisms is the perturbative decay of the inflaton field during its oscillation. During this perturbative reheating epoch, several physical phenomena might have occurred. However, there are currently very few probes available to unveil the physics occuring during this reheating phase, one of which is the inflationary 
tensor perturbation spectra. This spectra is highly sensitive to the 
equation of state (EOS) of the Universe during reheating. Given the upcoming era of gravitational wave (GW) astronomy, it is crucial to 
understand how the EOS of the Universe during the reheating era is 
influenced by the dynamics occurring at that time and how this, in turn, 
impacts the spectrum of inflationary tensor perturbations.

In this study, we highlighted that the shape of the inflationary potential 
during perturbative reheating era can affect the EOS of the Universe, 
thereby also influencing the spectrum of inflationary tensor perturbations. 
We have considered two different classes of the E-model $\alpha$-attractor 
inflationary potentials with $n = 1$ and $n = 3$ and studied the dynamics 
of the post-inflationary reheating era using two distinct approaches. In 
the first approach, the inflaton EOS is assumed to be a constant, while 
in the second approach, we account for the proper time evolution of the  
inflaton EOS. Beside determining the time evolution of the reheating EOS  
for various values of the inflaton decay rates, we also calculated how 
the propagation of primordial tensor perturbations is affected by this  
evolution of the reheating EOS. Our findings indicate that the 
inflationary tensor perturbations can be enhanced or suppressed by a 
factor of approximately 1.5-3, depending on the specifics of the  
inflationary potential and the inflaton decay rate assumed. Notably, 
for both the models, the present-day gravitational wave spectrum is 
expected to lie within the sensitivity range of the proposed resonance 
cavity experiments, making them likely to be measured with considerable 
accuracy. 

\appendix
\section{Effect of Primordial Tensor Perturbations on GW evolutions}
\label{sec:AppA}

It is clear from eq.~\ref{eq:gravspec0B} that the present-day GW spectrum 
is sensitive to the exact form of the inflationary tensor perturbation 
spectrum $\Delta^{2}_{k,{\rm inf}} (\equiv P_T)$ which is nearly 
scale-invariant around the pivot scale $k_*$ and is given by, 
$\Delta^{2}_{k,{\rm inf}} (\equiv P_T) \simeq 2H^2_{\rm inf}/\pi^2 M^2_{\rm Pl}$. However, for high-scale reheating scenarios like ours, the comoving 
modes re-enetering the horizon during reheating era corresponds to 
$k \gg k_*$, at which scale $P_T(k)$ is not strictly scale-invariant 
and a numerical estimation of $P_T(k)$ is necessary, as pointed out 
in~\cite{Garcia:2024rwg,Haque:2021dha}. In order to quantify the possible 
impact of this non scale-invariant $P_T(k)$ on our analysis, we numerically 
solve the tensor perturbation eq.~\ref{eq:GWevol1} during inflation and use 
the resulting $P_T(k)$ for evaluating the present-day GW spectrum 
$\Omega^0_{\rm GW}$ following eq.~\ref{eq:gravspec0B}.

\begin{figure}[t!]
\includegraphics[width=7.5cm,height=6.2cm]{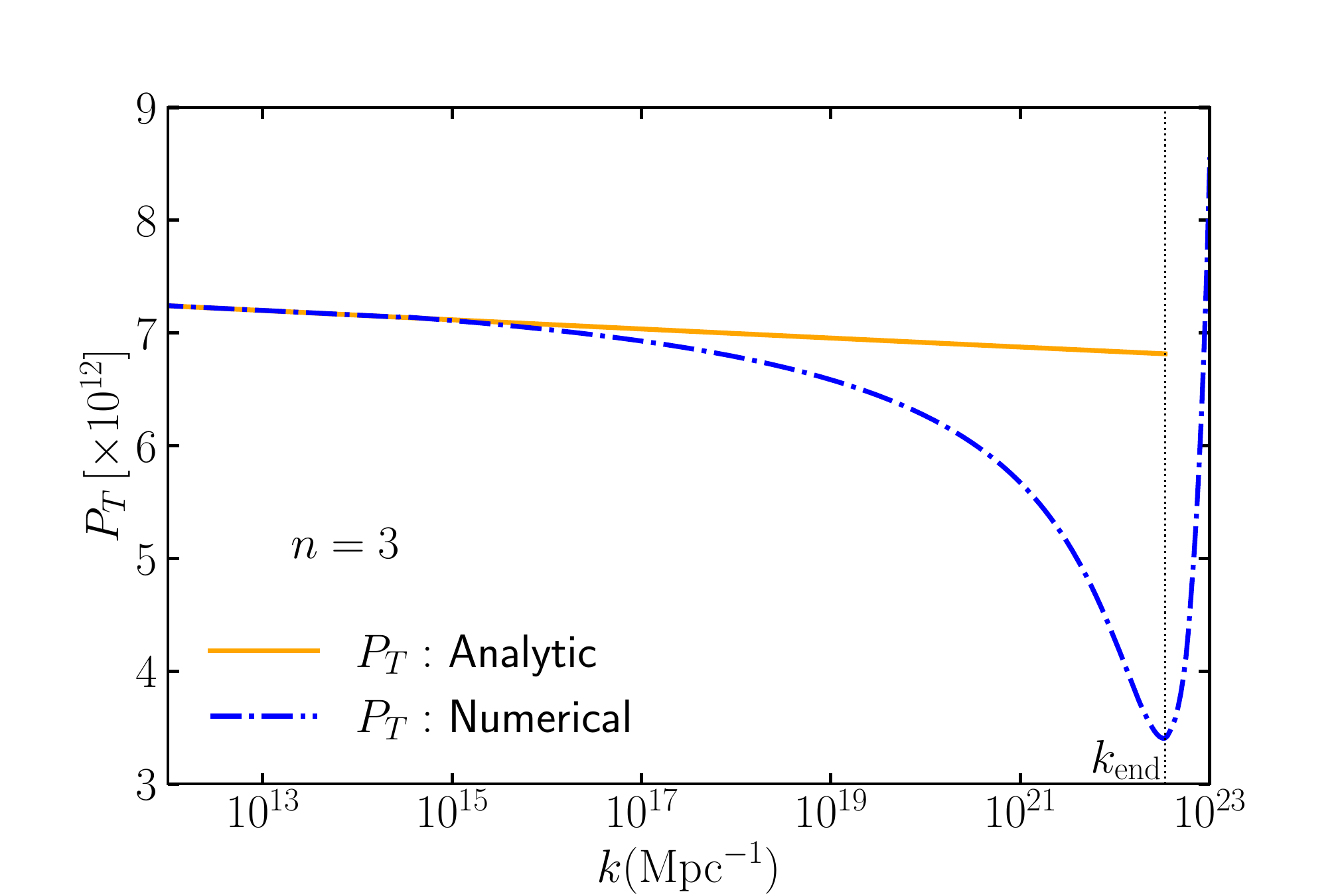}
\includegraphics[width=7.5cm,height=6.2cm]{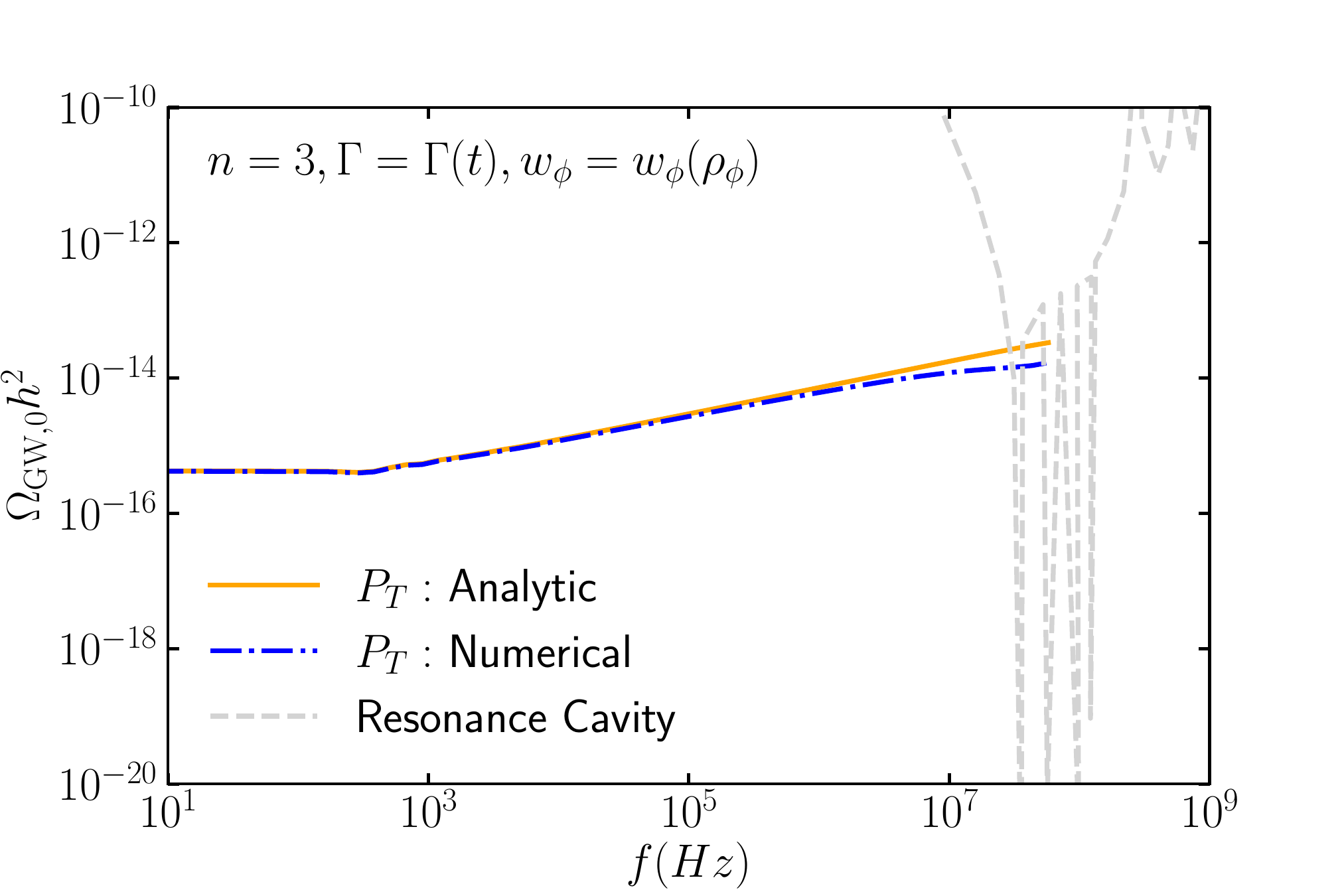}\\
\caption{{\it Left:} In case of $n=3$, a comparison of analytical vs numerical 
primordial tensor perturbation spectra ($P_T$) are shown. {\it Right:} 
Assuming time-dependent decay rate and actual inflaton EOS, the resulting 
present-day GW spectra are shown for $n=3$ case. See the text for details.}
\label{fig:PTvarspec}
\end{figure}

In the left panel of Fig.~\ref{fig:PTvarspec} we compare the analytical 
power-spectrum with the numerically obtained power-sepctrum for wide range 
of $k$ values in case of $n=3$ E-model $\alpha$-attractor potential, which 
shows that $P_T$ can be suppressed by a factor of $\sim 2$ for large values of 
$k$. In the left panel of Fig.~\ref{fig:PTvarspec} vertical black dotted 
line represents $k_{\rm end}$ which is the largest comoving mode exiting 
the horizon during inflation. Using this numerically obtained $P_T$ we 
also obtain the present-day GW spectrum considering the time-dependent 
decay rate of the inflaton using actual EOS of the inlfaton which is shown 
in the right panel of Fig.~\ref{fig:PTvarspec} by blue dashed-dotted line. 
Clearly, the GW spectrum is slightly suppressed at larger frequencies, 
compared to that obtained with analytic form of $P_T$ (shown by the 
orange line). However, such a suppression is also obtained if one considers 
the constant inflaton-EOS, and thus the impact of inflaton EOS on the 
present-day GW spectrum as presented in the main text remains unchanged.

\section*{Acknowledgment}
A.G. is funded by the ARC Centre of Excellence for Dark Matter Particle 
Physics, CE200100008. AG and DG thank Satyanarayan Mukhopadhyay for 
collaborating during the initial stage of the project.


\providecommand{\href}[2]{#2}\begingroup\raggedright\endgroup

\end{document}